\documentstyle[12pt]{article}
\catcode`\@=11
\def\be{%
\@ifnextchar[
{\def\ee{\end{equation}}\begin{equation}\l@b}%
{\def\ee{\]}\[}%
}
\def\bea{\begin{eqnarray}}
\def\eea{\end{eqnarray}}
\def\l@b[#1]{\label{#1}}

\def\eq#1{(\ref{#1})}
\def\Eq#1{Eq.(\ref{#1})}
\def\eqq#1{\hbox{eq.\eq{#1}}}
\def\Eqs#1-#2{Eqs.\eq{#1}--\eq{#2}}
\def\eqs#1-#2{\eq{#1}--\eq{#2}}
\catcode`\@=12
\begin{document}
%
%
\thispagestyle{empty}
\renewcommand{\thefootnote}{\fnsymbol{footnote}}
\vskip1.5in
\hfill \hbox{\vbox{\hbox{FERMILAB-PUB-92/164-T}
                   \hbox{PSU/TH/109}
                   \hbox{July 28, 1992}
             }}
\vskip2cm
\begin{center}
{\large
    BREAKDOWN OF DIMENSIONAL REGULARIZATION\\
[3mm]
    IN THE SUDAKOV PROBLEM.\\
[8mm]}
{J.~C.~Collins}%
\footnote{E-mail address: collins@phys.psu.edu.
}\\[2mm]
Physics Department, Penn State University,\\
104 Davey Lab., University Park, PA 16802, USA\\[3mm]
and \\[3mm]
{ F.~V.~Tkachov}%
\footnote{On leave from Institute for Nuclear Research of
the Russian Academy of Sciences, Moscow 117312, Russia;
supported in part by the Texas National Laboratory Research
Committee.
}\\[2mm]
Fermi National Accelerator Laboratory,\\
P.O.Box 500, Batavia, Illinois 60510 USA\\[2cm]
\end{center}
{
\centerline{\sc ABSTRACT}
\vskip2mm
\noindent
An explicit example is presented (a one-loop triangle graph)
where dimensional regularization
fails to regulate the infra-red singularities
that emerge at intermediate steps of studying
large-$Q^{2}$ Sudakov factorization.
The mathematical nature of the phenomenon is explained within the
framework of the theory of the $As$-operation.
}
\renewcommand{\thefootnote}{\arabic{footnote}}
\setcounter{footnote}{0}
\newpage
\thispagestyle{myheadings}
\paragraph{Introduction.}
The power of dimensional regularization \cite{dreg}
in applied Quantum Field Theory is nothing short of miraculous.
Its notational and ideological economy makes it a perfect tool
for a wide range of perturbative calculations%
\footnote{See, e.g., the record-setting 5-loop
   calculations in the $\varphi^{4}_{D=4}$ model \cite{5loop};
   the multiloop calculations in QCD \cite{QCDmultiloop};
   the industrial-scale QCD calculations of multijet
   hadronic scattering processes \cite{QCDcalc}.
\relax}
as well as a convenient instrument for theoretical studies of
factorization problems%
\footnote{See, e.g., the derivation of the general Euclidean
   asymptotic expansions of perturbative Green functions
   \cite{eu1},\cite{rest},\cite{rmp}.
\relax}.

Practical calculations in perturbative QCD typically exploit the
presence
of a large kinematic variable $Q$ and deal with (the leading terms
of)
the
corresponding asymptotic expansion of the cross section.  Results
describing the leading terms of such expansions generalize the
familiar Wilson short-distance operator product expansion and are
known as factorization theorems (for a review see e.g.
\cite{factor}).

The starting point of the construction of such an asymptotic
expansion is a
formal Taylor expansion of the integrand with respect to the
asymptotic parameter of the problem. Such an expansion typically
generates infrared singularities, in addition
to the ultraviolet divergences of the original
diagrams.  Even though there exist well-defined
prescriptions for eliminating the infrared singularities in
the (Wilson) coefficient functions of an expansion, the
singularities are important
at intermediate stages.

The main advantage of dimensional
regularization is that it simultaneously regulates both ultraviolet
and infrared divergences while preserving Lorentz and gauge
invariance. Situations where dimensional regularization works well
comprise all expansion problems of a Euclidean type
\cite{QCDmultiloop},\cite{eu1}
and many problems of an inherently non-Euclidean
(Minkowskian) nature \cite{QCDcalc}.

The aim of the present note is to show that there are singularities
that arise in the asymptotic analysis of Minkowski space problems
and that
cannot be regulated by dimensional continuation. We will pinpoint the
origin of the difficulties by means of an explicit one-loop example:
a
form factor graph. We will show that the nature of the problem is
very
general so that the same difficulty is bound to arise in many
Minkowski space situations.

A subsidiary aim is to popularize the concepts of a new and very
general method of analyzing asymptotic behavior of Feynman
graphs---the method of the $As$-operation for products of singular
functions \cite{eu1}.  The theory of the
$As$-operation has been worked
out in detail for the Euclidean problems \cite{eu1}, \cite{rest},
\cite{rmp},
where
it provides a simple and clean method for treating such problems as
renormalization, the operator product expansion, and the large-mass
expansion.  While the principles of $As$-operation \cite{eu1} are
very
general, there are several technical problems that one has to
overcome
before a fully satisfactory extension of the method to the most
general non-Euclidean situations is achieved. The results we describe
represent a first step in that direction.

\paragraph{The example.}
Consider the one-loop diagram of Fig.~1. We will see that neither the
possible presence of UV divergences nor the structure of the
numerator
of the corresponding integrand are relevant; only the structure of
the denominators is important. Therefore, it is irrelevant whether
the
corresponding particles are scalars, spinors or vectors. For
concreteness, one can imagine that the horizontal fat line is a
gluon in Feynman gauge while the side lines are quarks, and we will
use these names for the lines.
We consider the asymptotic behavior when
$Q^{2} \equiv -q^{2}=-(p_{1}-p_{2})^{2}$
gets large with all other invariants---$p_{1}^{2}$, $p_{2}^{2}$ and
the masses---held fixed.

To focus on the analytical effect we wish to describe, we will choose
the quarks to be massless, and the gluon to have mass $m$,%
\footnote{
Note that a non-zero gluon mass may have a dynamical origin
\cite{MassiveGluons}. Gauge invariance, however important for
studying
combinatorial properties of entire perturbation series, is of no
relevance to the analytical problem of asymptotic expansions of
individual graphs. If one wishes, one can consider the graph to be
one
in an abelian gauge theory, where the gluon can consistently be given
a mass.
\relax}
although the choice of masses does not affect the general
principles of our method.
The external quarks are on-shell: $p_{1}^{2}=p_{2}^{2}=0$.

We assume that the MS scheme \cite{MS} is used for UV
renormalization;
the dependence on the renormalization parameter $\mu $ thus
introduced
is logarithmic and known explicitly (the renormalized diagram has the
form of $\mu $-independent terms plus
${\rm constant}\times \log\mu $).
Anyhow,
since we will eventually concentrate on studying the {\em
integrand\/}
prior to loop integration, the UV behavior is also inessential.

There are two essential dimensional parameters in the problem:
$Q^{2}$
and $m^{2}$. The third parameter is the renormalization scale $\mu $.
The Sudakov asymptotic regime is $Q^{2}\to \infty $, with fixed
$m^{2}$ and $\mu $.  By dimensional analysis this is equivalent to
$m^{2}, \mu ^{2}\to 0$ with fixed $Q^{2}$.
(For definiteness, we will consider
the case of a space-like momentum transfer $q^{\mu }$.)
Since the dependence on $\mu $ is known explicitly,
it is sufficient to consider the expansion at $m^{2}\to 0$ with
$Q^{2}$
and $\mu $ fixed (cf.\ \cite{libby-sterman}). This is the most
convenient way to proceed within the techniques of the $As$-operation
\cite{eu1}.

The integrand for the graph is
\be[product]
   I(k;p_{1},p_{2},m) \equiv
     \frac {1}{ k^{2}-m^{2}  +i\eta }\times
     \frac {1}{ (k-p_{1})^{2}+i\eta }\times
     \frac {1}{ (k-p_{2})^{2}+i\eta }.
\ee
Considering contributions to the integral from various regions of
integration space is equivalent to considering various integrals of
the form
\be[integral]
     G[\phi ; p_{1},p_{2},m] \equiv
     \int d^{D}k \,I(k;p_{1},p_{2},m) \phi (k),
\ee
where $\phi $ is an arbitrary test function (i.e.\ a smooth function
which is non-zero only in a finite subregion of the integration
space).
So, one arrives at the basic problem of expanding arbitrary integrals
of the form \eq{integral} in $m\to 0$ with $p_{1}$ and $p_{2}$ fixed
and light-like. This is exactly the same as to say that one has to
expand the {\em integrand} \eq{product} in powers and logarithms of
the small parameter $m^{2}$ in the sense of distributions. Since such
expansions commute with multiplication by polynomials \cite{eu1}, one
can forget about numerators of propagators as well as possible vertex
factors.

Construction of an expansion in the sense of distributions involves
the following steps \cite{eu1}: formal Taylor expansion of the
product
\eq{product} in powers of $m$; classification of singularities of
such a
formal expansion; construction of counterterms to be added to the
formal expansion to transform it into a correct expansion in the
sense
of distributions. Construction of counterterms proceeds in an
iterative fashion: from singularities of simpler nature---i.e., to
which fewer singular factors contribute---to the more complicated
ones; construction of counterterms for the latter involves simpler
counterterms obtained at previous iterations.

For practical reasons
it
is convenient, whenever possible, to employ a regularization
throughout the entire procedure.  Dimensional regularization is
the
prime candidate for that role. However, as we will see shortly, it
fails to regulate a singular expression at an intermediate step of
constructing the expansion in the sense of distributions for
\eq{product}.

\paragraph{Geometry of singularities of the formal expansion.}
We follow the general procedure that gives the $As$-operation. First
we make a formal expansion of the integrand in powers of $m$.  This
gives a correct leading order expansion of the integral with a test
function, provided that the test function is zero in a neighborhood
of
all singularities of the expanded integrand.  That is
\bea
     G[\phi ; p_{1},p_{2},m] &=&
     \int d^{D}k \, \phi (k) \,
     \frac {1}{k^{2}        +i\eta  }\times
     \frac {1}{(k-p_{1})^{2}+i\eta  }\times
     \frac {1}{(k-p_{2})^{2}+i\eta  }
\nonumber \\
  && \!\! \mbox{} + o(1) ,
     \qquad \hbox{if $\phi $ is zero near singularities} .
\label{nonsing}
\eea
Here the remainder is a power of $m$ smaller than the leading term.

To get a complete expansion, valid for all test functions, one should
examine a small neighborhood of the singularities of the first term
on
the right of \Eq{nonsing}. The principles of the $As$-operation
prescribe
that we should start with the highest dimension singular surface and
then treat successively lower dimension surfaces.

Since the integration contours may be deformed away from the
light-cone singularity of any single propagator (except at its
apex), such a singularity is integrable.  Hence the only
singularities that we need be concerned with are the apexes of
the light cones and the intersections of
the light-cone singularities of different propagators,

Now, the first denominator in \Eq{nonsing} generates a singularity
localized on the light-cone $k^{2}=0$. The second factor is singular
on $(k-p_{1})^{2}=0$ which is nothing but the light-cone $k^{2}=0$
shifted so that its apex is at the point $k=p_{1}$.
Since $p_{1}$ is a light-like vector, the two light cones intersect
on
the straight line
\be[line1]
     A=\{k=z_{1}p_{1}, \quad -\infty <z_{1}<+\infty \} .
\ee
This intersection is non-trivial because the light cones are not
transverse at the intersection points.
Similarly, the singularities of the first and third factor overlap
on the line
\be[line2]
     B=\{k=z_{2}p_{2}, \quad -\infty <z_{2}<+\infty \} .
\ee

The singularities of the second and third factors intersect on a
smooth manifold and are transverse there; this implies that in a
small
neighborhood of each intersection point the integral factorizes,
so that the singularity is integrable and
such an intersection is harmless.
The exception is the point $k=0$ where the first
denominator is also zero, and we consider this point separately.

There are three points---namely, $k=0$, $k=p_{1}$, and
$k=p_{2}$---where the analytical nature of singularities is more
complicated than at the generic points of the lines
\eqs{line1}-{line2}. In particular, the singularity
\be[S]
     S=\{k=0\},
\ee
is where the effect we are after takes place.

The geometrical pattern of singularities can be visualized as in
Fig.~2.

\paragraph{Structure of singularities at $k\propto p_{1}$.}
Consider singularities near a generic point on the line $A$,
\eq{line1}. Fix $z\neq0,1$, and consider a small neighborhood $\cal
O$
of the point $zp_{1}$. The third factor of the integrand is smooth in
$\cal O$ and, therefore, can be effectively relegated to the test
function. So, within $\cal O$ it is sufficient to study the expansion
of the product of only the first two factors that contribute
non-trivially to the singularity in $\cal O$. It is convenient to
choose light-cone coordinates, $k=(k_{+},k_{-},k_{\perp })$, so that
$p_{1}=(p_{1{+}},0,0_{\perp })$,
$p_{2}=(0,p_{2{-}},0_{\perp })$.
Our conventions will be such that
$k^{2}=k_{+}k_{-}-k_{\perp }^{2}$,
$2k\cdot p_{1}=p_{1{+}}k_{-}$,
and $d^{D}k=\frac {1}{2}dk_{{+}}dk_{{-}}d^{D-2}k_{\perp }$.

Then the formal expansion of the first two factors takes the form:
\be
     \frac {1}{z    p_{1{+}}k_{-}-k_{\perp }^{2} -m^{2}+i\eta }\times
     \frac {1}{(z-1)p_{1{+}}k_{-}-k_{\perp }^{2}       +i\eta }
\ee\be[two]
   =
     \frac {1}{z    p_{1{+}}k_{-}-k_{\perp }^{2} +i\eta }\times
     \frac {1}{(z-1)p_{1{+}}k_{-}-k_{\perp }^{2} +i\eta }
     + o(1).
\ee
One can see that the singularities of the product on the r.h.s.\ are
localized at the origin of the space of the variables $k_{-}$ and
$k_{\perp }$. This means that \eq{two} holds in the sense of
distributions on test functions that are zero near $k_{-}=k_{\perp
}=0$.
The situation here is very similar to what one has in the case of a
single Euclidean propagator treated in section 7 of \cite{eu1}---see
especially eqs.~(7.25) and (7.26).  The only important difference is
that the singular functions on the r.h.s.\ are not homogeneous.  This
is remedied by the change of variable $k_{-}=\pm t^{2}$ after which
the expressions on the r.h.s.\ of \eq{two} become homogeneous with
respect to simultaneous scaling in $t$ and $k_{\perp }$, and simple
power counting shows that the singularity is logarithmic (at $D=4$).

This allows one to repeat the reasoning of section 7 of \cite{eu1} to
write down the following analogue of eq.~(7.25) of \cite{eu1}:
\bea
     \hbox{The l.h.s.\ of \eqq{two}}
   &=&
     \frac {1}{z    p_{1{+}}k_{-}-k_{\perp }^{2} +i\eta }\times
     \frac {1}{(z-1)p_{1{+}}k_{-}-k_{\perp }^{2} +i\eta }
\nonumber\\
\label{two-expan-k}
   && \mbox{} +
     c_{A}(m^{2},z) \frac {1}{p_{1+}}
     \delta (k_{-})\delta ^{(D-2)}(k_{\perp })
     + o(1),
\eea
with
\bea
     c_{A}(m^{2},z)
     &=&
     p_{1{+}}\int dk_{-}d^{D-2}k_{\perp }\,
     \left[
        \frac {1}{z    p_{1{+}}k_{-}-k_{\perp }^{2}-m^{2}+i\eta
}\times
        \frac {1}{(z-1)p_{1{+}}k_{-}-k_{\perp }^{2}      +i\eta }
     \right.
\nonumber \\
   && \hspace{1.5in}
      \left. \mbox{} -
        \frac {1}{z    p_{1{+}}k_{-}-k_{\perp }^{2} +i\eta }\times
        \frac {1}{(z-1)p_{1{+}}k_{-}-k_{\perp }^{2} +i\eta }
     \right]
\nonumber\\
\label{ct-k}
     &=&
     p_{1{+}}\int dk_{-}d^{D-2}k_{\perp }\,
        \frac {1}{z    p_{1{+}}k_{-}-k_{\perp }^{2}-m^{2}+i\eta
}\times
        \frac {1}{(z-1)p_{1{+}}k_{-}-k_{\perp }^{2}      +i\eta } .
\eea
In the second form of this integral we have used the fact that
the integral of a homogeneous function\footnote{
Recall that the reasoning that leads to this expression is,
strictly speaking, done in the ``straightened'' coordinates
$(t,k_{\perp })$ with homogeneous functions.
\relax}
is zero within dimensional regularization.

At this point dimensional
regularization works.
Recall \cite{eu1} that the role of the counterterm $c_{A}$ is
two-fold: first, it cancels the divergence of the first singular
expression (after integration with test functions); second, it
ensures that the expansion is asymptotic ($o(1)$) on any test
function.
The expansion \eqs{two-expan-k}-{ct-k} holds in the sense of
distributions on any test function (in the variables
$k_{-},k_{\perp }$).
The simplification gained is that, whereas the l.h.s.\ of
\eq{two-expan-k} has a complicated dependence on $m$, the
first two terms on the right have a simple power dependence,
which becomes logarithmic at $D=4$.  This we will verify from the
explicit calculation of $C_{A}$ in the next section.

\paragraph{Explicit expressions for counterterms.}
\label{expl}
It is not difficult to perform the integrations in \eq{ct-k}
explicitly%
\footnote{
Recall that the definition of dimensional regularization \cite{dreg}
prescribes that we should perform the integrations over the
transverse components first.
In the present case, however, the same result is reproduced in a
slightly easier way if one first performs integration over $k_{-}$.
At
this point one should be aware of the fact that the underlying
definition of the integral is in terms of the homogeneous
coordinates,
and the cutoffs that are present at intermediate stages (cf.\
\cite{eu1}) are symmetric with respect to the homogeneous coordinates
and become asymmetric in the coordinates $k_{-}, k_{\perp }$. Such
subtleties stress the need for a meaningful rigorous definition of
dimensional regularization in momentum/coordinate space
representations. Some of the issues involved will be discussed in a
forthcoming publication \cite{dregf}.
\relax}
to obtain
\be[ct-expl]
     c_{A}(m^{2},z)
     =
     2i \, \theta (0<z<1)\,\Gamma (\epsilon )\,
     \pi ^{2-\epsilon }
     \left[(1-z)m^{2}\right]^{-\epsilon },
\ee
where $\epsilon =(4-D)/2$.

There are a few points worth making here. First, the counterterm
is zero for $z>1$ and $z<0$. This agrees with the fact that
according to the Landau equations%
\footnote{
See e.g.\ \cite{landauEq}. Note that the Landau equations are usually
associated with studying analyticity properties of Feynman diagrams.
Finding cuts etc., however, is equivalent to determining when an
expansion near the corresponding value of external momenta and masses
contains non-trivial (i.e.\ non-analytic) contributions. In the
context of the expansion problem proper this issue was reconsidered
by
Libby and Sterman \cite{libby-sterman}. A reinterpretation of the
Landau equations from the point of view of asymptotic expansions of
distributions is presented in \cite{FVTlandauEq}.
\relax}
the singularities at those values of $z$ are not pinched. In the
language of distribution theory, this says that
for $z<0$ and $z>1$ the product in the
first term of the r.h.s.\ \eq{two-expan-k} is ($i$)
well-defined in the sense of distributions and ($ii$) does not
require
additional counterterms to approximate the l.h.s.\ in the sense of
distributions to $o(1)$.  (In general, ($ii$) does not necessarily
follow from ($i$).)

Second, the counterterm \eq{ct-expl} is the only term in
\eq{two-expan-k} that contains a non-analytic dependence on $m$
(cf.\ the discussion of the role of such counterterms in \cite{eu1}).
After expansion in $\epsilon $ the dependence becomes logarithmic.%
\footnote{
   It is possible to avoid the use of dimensional regularization
   in expressions like \eqs{two-expan-k}-{ct-k}---cf.\
   \cite{rmp} where the results of \cite{eu1} are presented in a
   regularization-independent form.
\relax}

Third, the expression \eq{ct-expl} and all its derivatives in $z$
have
well-defined limits as $z\to +0$. This conclusion does not
change if the second (quark) propagator in \eq{product} and,
correspondingly, in the first term of \eq{two} contains a non-zero
mass, say, $m_{1}$, because then the last factor in \eq{ct-expl} will
simply be replaced by
$\left[zm_{1}^{2}+(1-z)m^{2}\right]^{-\epsilon }$.
Note, however, that the presence of mass in the first factor (the
bottom line or gluon in Fig.~1) {\em is\/} important.

Finally, it is not difficult to rewrite the above expansion in an
explicitly covariant form:
\bea
\lefteqn{
     \frac {1}{k^{2}-m^{2}  +i\eta }\times
     \frac {1}{(k-p_{1})^{2}+i\eta }
}
\nonumber \\
\label{two-expan-cov}
 \qquad  &=&
     \frac {1}{k^{2}      +i\eta }\times
     \frac {1}{(k-p_{1})^{2}+i\eta }
     +
     \int _{0}^{1} dz\,\delta ^{(D)}(k-zp_{1})\,c_{A}(m^{2},z)
     +
     o(1).
\eea
We have explicitly taken into account that $c_{A}$ vanishes outside
the interval $0<z<1$. Note that the $\delta $-function on the r.h.s.\
is $D$-dimensional. This expression represents the leading power term
in a correct asymptotic expansion in the sense of distributions on
test functions that are zero in small neighborhoods of $k=0$ and
$k=p_{1}$.
Note that a similar expansion is obtained for the expansion
of the product of the first and third factors of \eq{product}
(with $p_{1}$ replaced by $p_{2}$).

\paragraph{Taking into account the third factor.}
Let us now consider the entire expression \eq{product}.  In order to
transform its formal expansion, the integrand of \eq{nonsing}, into a
well-defined expansion in the sense of distributions, the general
recipe of the Extension Principle of \cite{eu1} tells to add
counterterms localized at singular points of the formal expansion in
\eq{nonsing} with properly chosen coefficients.
Let us show that the expansion that is valid on the test functions
that vanish in neighborhoods of the points $k=0$, $k=p_{1}$ and
$k=p_{2}$
is given by the following formula:
\bea
   \lefteqn{
     \frac {1}{k^{2}-m^{2}  +i\eta }\times
     \frac {1}{(k-p_{1})^{2}+i\eta }\times
     \frac {1}{(k-p_{2})^{2}+i\eta }
}
\nonumber\\
  \qquad &=&
     \frac {1}{k^{2}        +i\eta }\times
     \frac {1}{(k-p_{1})^{2}+i\eta }\times
     \frac {1}{(k-p_{2})^{2}+i\eta }
\nonumber\\
   && \mbox{} +
     \frac {1}{(k-p_{2})^{2}+i\eta }\times
     \int _{0}^{1} dz\,\delta ^{(D)}(k-zp_{1})\,c_{A}(m^{2},z)
\nonumber\\
\label{three+ct}
   && \mbox{} +
     \frac {1}{(k-p_{1})^{2}+i\eta }\times
     \int _{0}^{1} dz\,\delta ^{(D)}(k-zp_{2})\,c_{B}(m^{2},z)
     +
     o(1).
\eea
Indeed, any test function $\varphi(k)$ that vanishes in small
neighborhoods of the points $k=0$, $k=p_{1}$ and $k=p_{2}$, can be
represented as a sum $\varphi_{1}+\varphi_{2}$ where $\varphi_{1}$
is zero around the line $k\propto p_{2}$ and $\varphi_{2}$ is zero
around the line $k\propto p_{1}$. On $\varphi_{1}$, the second
counterterm
vanishes and we are left with the product of
the expansion \eq{two-expan-cov} times the third factor.
Since the product of the third factor and $\varphi_{1}$ is a valid
test function $\tilde\varphi_{1}$, one arrives at a correct expansion
to $o(1)$. A similar reasoning is applied to $\varphi_{2}$ (note that
the function $c_{B}$ coincides with $c_{A}$ in our example).

It follows that the expansion \eq{three+ct} is actually
correct for all test functions that vanish in small
neighborhoods of the points $k=0$, $k=p_{1}$ and $k=p_{2}$.
To make the expansion valid on {\it all\/} test functions,
one has to add appropriate
counterterms localized at the points $k=0$, $k=p_{1}$, and $k=p_{2}$.
Such counterterms are, in general, linear combinations of
$\delta $-functions and their derivatives localized at those points
with coefficients depending on the expansion parameter
in a non-analytic (logarithmic) manner \cite{eu1}.

To construct the additional counterterms, one
must ($i$) perform an
appropriate power counting in order to
determine the strength of the singularity;
($ii$) introduce an
intermediate regularization to make the singularity manageable (in
Euclidean problems \cite{eu1} dimensional regularization
automatically
regulates all singularities) or perform an explicit subtraction (as
in \cite{rmp});
($iii$) determine an explicit form of the
counterterms that need to be added in order to ensure the
approximation property of the resulting expansion.

It will be at step ($ii$) that dimensional regularization fails
in our example.

\paragraph {Singularity at $k=0$.}
Let us focus on the point $k=0$. One has to study the singularity
of the entire r.h.s.\ of \eq{three+ct} at $k\to 0$.
The r.h.s.\ of \eq{three+ct} contains contributions
that are analytic in $m$, and those that are not.
It is sufficient to consider the latter since they cannot be
affected by how the analytic contributions are treated, and it is
the non-analytic terms that will exhibit the failure of dimensional
regularization.

The terms with non-analytic dependence in $m$ are known explicitly in
our case:
\bea
     \left[
     \frac {1}{(k-p_{2})^{2}+i\eta }\times
     \int _{0}^{1} dz\,\delta ^{(D)}(k-zp_{1})\,c_{A}(m^{2},z)
     +
     (1 \leftrightarrow 2)
     \right]
\nonumber\\
\label{nonan}
  =
     \left[
     \frac {1}{Q^{2}}
     \int _{0}^{1} dz\,
     \frac {1}{z-i\eta }\,
     \delta ^{(D)}(k-zp_{1})\,c_{A}(m^{2},z)
     +
     (1 \leftrightarrow 2)
     \right] .
\eea
One can immediately see that:
\begin{itemize}
\item[($a$)] One has to deal with the product of a
one-dimensional distribution $1/(z-i\eta )$, which itself
is well-defined if integrated with {\em smooth\/} test functions,
times $\theta (z)$.  The resulting expression is
singular and ill-defined at $z=0$.  The distribution $1/(z-i\eta )$
is generated from the propagator $1/[(k-p_{2})^{2}+i\eta ]$ when we
set
$k=zp_{1}$.
\item[($b$)] Dimensional regularization does not regulate this
singularity
because the form of the product is independent of $\epsilon $ and
there remain no ``unused" extra dimensions.
\item[($c$)] There are no cancellations at $k=0$
between contributions from the two counterterms corresponding
to the two singular lines.
\end{itemize}

It remains to note that the effect of breakdown of dimensional
regularization in the above example persists
(even if the formulas become more cumbersome) if one introduces
masses
of order $m$---say, $m_{1}$ and $m_{2}$---into the two quark
propagators,
or allows the external quarks to be off-shell by $O(m^{2})$.
One can also see that the same configuration of singularities
emerges e.g.\ in the studies of the large-$s$ limit (see Fig.~3).
All this points to universality of the phenomenon of the breakdown
of dimensional regularization in expansion problems in
non-Euclidean asymptotic regimes.

\paragraph{Conclusions.}
We have considered a rather typical non-Euclidean expansion
problem (a one-loop form factor graph in the Sudakov asymptotic
regime)
within the framework of the theory of $As$-operation,
and we saw some significant differences from Euclidean problems.
In particular, we have discovered a class of singularities which
are not regulated by dimensional regularization.

The origin of the dimensionally-nonregularizable singularities
in non-Euclidean asymptotic expansion problems
is completely general:

{\bf First},
the non-trivial (``pinched") singularities of the expanded
integrands in the case of non-Euclidean asymptotic regimes may be
localized on manifolds with boundaries (which is never the case for
Euclidean regimes, where singularities are always localized on linear
subspaces of the space of integration momenta).

{\bf Second,}
construction of a complete expansion requires introduction of
counterterms that contain the non-analytic dependence
on the expansion
parameter and are localized on such manifolds;
such counterterms may have
coefficients that, together with their derivatives, possess finite
non-zero limiting values at the boundaries of such manifolds when the
boundaries are approached from within the manifold ($z\to +0$ in our
case)
while being identically zero outside the boundary. In other words,
if the boundary is described by the equation $z=0$ in local
coordinates
with $z>0$ corresponding to the pinched submanifold,
then the coefficients near the boundary have the form
$\theta (z)\times $
(a smooth function of $z$)---even for $D\neq4$.

{\bf Third,}
light-cone singularities of the factors that do not contribute
to such counterterms may pass over the boundaries of the
corresponding
manifolds; when projected onto such manifolds (which is exactly what
happens when one introduces the counterterms into the entire
product---cf.\ \eq{nonan}), they take the form $1/({z\pm i\eta })$.

{\bf Fourth,}
such manifolds may be geometrically positioned so
that the extra dimensions
that are instrumental in the mechanism of dimensional regularization
are ``used up" in the counterterms and do not provide
any suppression for the resulting singular product of the type
$\theta (z)\times 1/({z\pm i\eta })$.

It should be emphasized that
the problem here is not an ambiguity as in the case of $\gamma _{5}$,
but a failure of dimensional regularization to regulate
a particular class of infra-red singularities.
Moreover, since expansions in the sense of distributions
in powers and logarithms of the expansion parameter are
unique, one has to conclude that it is impossible to get rid of the
problem by choosing a different ``factorization scheme".

The problem is certainly associated with our insistence on strict
power-and-logarithm
expansions: The $1/z$ singularity gives a problem
because we have expanded everything else in the integrand in
powers of a small variable. However, the requirement
that the expansions to be constructed run in powers and logarithms of
the expansion parameter
(the requirement of ``perfect factorization" \cite{eu1})
cannot be relaxed for both phenomenological and technical reasons.
In particular, such expansions possess
the property of uniqueness which greatly facilitates
iterative construction of the expansions, relieving one of having
to worry about unitarity and gauge invariance of the final results
etc.\ \cite{eu1}.  This is particularly true
in non-Euclidean problems, since it is only at the leading
twist level that we get simple factorization theorems.%
\footnote{In the method one of us has given \cite{JCCSudakov} for
   treating the Sudakov form factor at the complete leading twist
   level, non-dimensionally-regularizable singularities are
   avoided either by the use of axial gauge or by the use of an
   equivalent trick in Feynman gauge.  One consequence of the
   resulting lack of ``perfect factorization'' is an annoying
   proliferation of remainder terms in Ward identities.  These
   are especially tricky to handle in a non-abelian theory. }

On the other hand, the existence and nature of the
anomalies one may have to deal with as a result of
the effect we have described is not obvious.
One thing is clear, however: whether one opts for
other regularizations (e.g.\ analytic regularization which would
replace $1/({z\pm i\eta })$ in the above expressions by
$(z\pm i\eta )^{-\lambda }$),
or chooses to combine dimensional regularization with
a formalism involving direct subtractions as in \cite{rmp}---or
to forgo dimensional regularization altogether in favor of the
latter---the consequences may be rather unpleasant both for practical
calculations and for the general theory of higher-twist
factorization.

\vskip1cm
\noindent {\em Acknowledgments.}
One of the authors (F.T.) is grateful for hospitality to the Physics
Department of Penn State University and to the Theory Division of
Fermilab where parts of this work were done,
and to A.~V.~Radyushkin for an illuminating discussion.
This work was supported in part by the U.S. Department of Energy
under grant DE-FG02-90ER-40577, and by the Texas National
Laboratory Research Commission, as part of the CTEQ collaboration.

%
%
%
\newpage\thispagestyle{myheadings}\markright{}

\newpage
\thispagestyle{myheadings}\markright{}
\centerline{Figure captions.}
\begin{itemize}
%
%
%
\item[Fig.~1.] The triangle graph corresponding to our example
\eq{product}.
The bottom line corresponds to a gluon with non-zero mass $m$,
and the side lines to massless quarks.
%
%
%
\item[Fig.~2.] The geometrical pattern of singularities due to the
three
denominators of the formal expansion, \eq{nonsing}.
%
%
%
%
\item[Fig.~3.] A configuration of propagators (and singularities)
essentially similar to that in
Fig.~1 emerges in the large-$s$ small-$t$ problem.
\end{itemize}
%

\begin{thebibliography}{99}

\bibitem{dreg}
G.~'t Hooft and M.~Veltman: Nucl.~Phys.~B44 (1972) 189;
for a review see e.g.\ \cite{collins}.
\bibitem{collins}
J.~C.~Collins: ``Renormalization'', Cambridge
University Press, 1984.
\bibitem{5loop}
F.~V.~Tkachov: Phys.~Lett.~100B (1981) 65;
K.~G.~Chetyrkin, S.~G.~Gorishny, S.~A.~Larin and F.~V.~Tkachov:
Phys.\ Lett.\ 132B (1983) 351;
S.~G.~Gorishny, S.~A.~Larin and F.~V.~Tkachov:
Phys.~Lett.~101A (1984) 120.
\bibitem{QCDmultiloop}
O.~V.~Tarasov, A.~A.~Vladimirov and A.~Yu.~Zharkov:
Phys.~Lett.~93B (1980) 429;
S.~A.~Larin, F.~V.~Tkachov, and J.~A.~M.~Vermaseren:
Phys.~Rev.~Lett.~66  (1991) 862;
S.~G.~Gorishny, A.~L.~Kataev and S.~A.~Larin:
Phys.~Lett.~259B (1991) 144.
\bibitem{QCDcalc}
R.~K.~Ellis, M.~A.~Furman, H.~E.~Haber and I.~Hinchliffe:
Nucl.~Phys.~B173 (1980) 397;
T.~Matsuura, S.~van der Marck and W.~van Neerven:
Nucl.~Phys.~B319 (1989) 570;
W.~T.~Giele and E.~W.~N.~Glover:
FERMILAB-PUB-91/100-T (to be published in Phys.~Rev.~D).
\bibitem{factor}
J.~C.~Collins, D.~E.~Soper and G.~Sterman:
``Factorization of hard processes in QCD'',
in: ``Perturbative QCD'' (A.~H.~Mueller, ed.),
World Scientific, Singapore, 1989.
\bibitem{eu1}
F.~V.~Tkachov:
``Euclidean Asymptotic Expansions of Green Functions of Quantum
Fields.
(I) Expansions of Products of Singular Functions",
preprint FERMILAB-PUB-91/347-T;
Int.~J.~Mod.~Phys.~A (1992)
(in~print)~and~refs.\ therein.
\bibitem{rest}
G.~B.~Pivovarov and F.~V.~Tkachov:
``Euclidean Asymptotic
%
%
Expansions of Green Functions of Quantum Fields.
(II) Combinatorics of the
$As$-operation", preprint FERMILAB-PUB-91/345-T;
Int.~J.~Mod.~Phys.~A~(1992)~(in~print);.
%
\bibitem{rmp}
A.~N.~Kuznetsov, F.~V.~Tkachov and V.~V.~Vlasov:
``Techniques of Distributions in
Perturbative Quantum Field Theory.
(I) $As$-operation for products of singular functions",
preprint PSU/TH/108, June 1992 (to be published).
%
\bibitem{MassiveGluons}
J.~M.~Cornwall: Phys.~Rev.~D26 (1982) 1453.
\bibitem{MS}
G.~'t~Hooft: Nucl.~Phys.~B61 (1973) 455.
\bibitem{landauEq}
R.~J.~Eden, P.~V.~Landshoff, D.~I.~Olive and J.~C.~Polkinghorne:
``The Analytic S-Matrix", Cambridge University Press, 1966.
\bibitem{libby-sterman}
S.~Libby and G.~Sterman: Phys.~Rev.~D18 (1978) 3252, 4737.
\bibitem{FVTlandauEq}
F.~V.~Tkachov: in preparation.
%
\bibitem{dregf}
A.~N.~Kuznetsov and F.~V.~Tkachov: (in preparation).
%
\bibitem{JCCSudakov}
J.~C.~Collins, Phys.~Rev.~D22 (1980) 1478; ``Sudakov Form
Factors''
in: ``Perturbative QCD'' (A.~H.~Mueller, ed.),
World Scientific, Singapore, 1989.

\end{thebibliography}
\end{document}